\documentclass[twocolumn]{aastex631}

\usepackage{graphicx}
\usepackage[none]{hyphenat}
\usepackage{amsmath}
\usepackage{booktabs}
\usepackage{multirow}
\usepackage{txfonts}
\usepackage{enumitem}
\usepackage{color}
\usepackage{ulem}
\usepackage{threeparttable}
\usepackage{algorithmic}
\usepackage{bm}
\usepackage{mathrsfs}
\usepackage{booktabs}  
\usepackage{float}
\usepackage{subfigure}
\DeclareRobustCommand{\VAN}[3]{#2}
\let\VANthebibliography\thebibliography
\def\thebibliography{\DeclareRobustCommand{\VAN}[3]{##3}\VANthebibliography}
\usepackage{color}

\usepackage[mathlines]{lineno}

\begin{document}

\title{Influence of Density Distribution on Synchrotron Polarization Dispersion in Magnetized Interstellar Medium }

\author{Ya-Wen Xiao}
\affiliation{\rm{Department of Physics, Xiangtan University, Xiangtan, Hunan 411105, People’s Republic of China};\\}
 
\author{Jian-Fu Zhang}
\affiliation{\rm{Department of Physics, Xiangtan University, Xiangtan, Hunan 411105, People’s Republic of China};\\}\email{jfzhang@xtu.edu.cn}

\author{Alex Lazarian}
\affiliation{\rm{Department of Astronomy, University of Wisconsin, Madison, WI, USA}}

\author{Dmitri Pogosyan}
\affiliation{\rm{Physics Department, University of Alberta, Edmonton, Canada}}

\begin{abstract}
Faraday rotation measure (RM) synthesis is a well-known approach originated in Burn (1966) and later developed by Brentjens \& de Bruyn (2005) for studying magnetic fields. This work presents a complementary approach--the polarization frequency analysis (PFA)--allowing for the properties of the turbulent magnetic field, which are difficult to include in Burn's original approach. Based on synthetic polarization observation of magnetohydrodynamic turbulence simulation data, we study the influence of the coupling effect between density and magnetic field on synchrotron polarization dispersion. By applying the PFA to different simulated interstellar turbulence environments, we find that the PFA technique can reveal the scaling slope of the turbulent magnetic field in the case of a weak coupling effect and can also reflect the scaling slope of the RM in the case of a strong coupling effect. Since it avoids the influence of Faraday depolarization, the PFA technique is a promising way to uncover turbulence properties using observational data from the Low-Frequency Array for Radio Astronomy and the Square Kilometre Array.
\end{abstract}


\section{Introduction} \label{Sect_intro}
Currently, astronomical observational evidence indicates that the interstellar medium (ISM) is turbulent and magnetized, e.g., the big power law of electron density (\citealt{Armstrong1995ApJ, CL2010ApJ}), fluctuations of velocity in the solar wind (\citealt{Leamon1998JGR, Bruno2005LRSP}) and observation of the synchrotron emission (\citealt{McClure-Griffiths2001ApJ, Gaensler2001ApJ, Iacobelli2013A&A}). A large number of key astrophysical processes are related to magnetohydrodynamic (MHD) turbulence, such as the diffusion and acceleration of cosmic rays (\citealt{Yan2008ApJ, Zhang2021ApJ, Gao2024ApJ, Gao2025A&A}), heat conduction (\citealt{Narayan2001ApJ, Lazarian2006ApJ}), turbulent magnetic reconnection (\citealt{LV99}), interstellar gas heating (\citealt{Phillips2023MNRAS}), ionization of plasmas (\citealt{Forteza2007A&A}) and the formation of stars (\citealt{MacLow2004, McKee2007ARA&A}). The mechanism by which turbulence is driven remains an open question. In general, external driving, such as supernova explosions (\citealt{Bacchini2020A&A}) and merger events (\citealt{Subramanian2006MNRAS}), is an acceptable way. Besides, various instabilities such as tearing mode, Kelvin-Helmholtz, and Rayleigh-Taylor instabilities (\citealt{Furth1963PhFl, Reynolds2005MNRAS, Kowal2020ApJ}) are also considered to be one of the potential ways of driving turbulence. An observationally oriented understanding of the nature of turbulence will help deepen the application of MHD turbulence in the key astrophysical environments mentioned above.

When exploring the magnetic field and velocity, we need to remove the influence of density fluctuations that impede the measurement of magnetic field and velocity (\citealt{L1995A&A}). For instance, velocity channel analysis and velocity coordinate spectrum using spectroscopic observational data (\citealt{LP2000ApJ, LP2006ApJ, Lazarian2009}) have been developed to decouple the contributions of turbulent densities and velocities (see \citealt{Yuen2021ApJ} for decoupling the velocity and density). Note that the presence of shock waves in molecular clouds results in a clump density following a log-normal distribution, with a shallower scaling index than the Kolmogorov one (\citealt{BLC2005ApJ, Kowal2007ApJ}). After correctly understanding the properties of density fluctuations, we can determine the properties of velocity.

Relativistic electrons diffused in the ISM interacting with magnetic fields inevitably produce synchrotron emission fluctuations, which is considered the theoretical basis for developing analysis techniques to trace magnetic field information. Using fluctuation statistics of synchrotron intensity (\citealt{LP12ApJ}) and polarization intensity (\citealt{ LP2016ApJ}; hereafter LP16) can reveal the properties of the magnetic field perpendicular to the line of sight (LOS). When considering the Faraday rotation effect, statistical analysis of synchrotron polarization intensity can reflect 3D magnetic field information (LP16; \citealt{Lazarian&Yuen2018ApJ}). However, Faraday depolarization will be involved in the effects of density. Therefore, the properties of density should determined first (e.g., see \citealt{XZ2020ApJ} for using the dispersion measure method), and then one can better extract the properties of the magnetic field. In addition, the analysis of synchrotron polarization intensity has also been suggested to explore the other properties of MHD turbulence cascade such as the anisotropy (\citealt{Herron2016ApJ, Wang2020ApJ}), the direction of the magnetic field (\citealt{Lazarian&Yuen2018ApJ, Zhang2019MNRAS, zhang2019ApJ}), scaling slopes (\citealt{Zhang2016ApJ, Zhang2018ApJ, Zhang2025}), and magnetization (\citealt{Carmo2020ApJ, Guo2024ApJ}; {see also \citealt{Zhang2022FrASS} for a review}).

Similar to the velocity channel analysis and velocity coordinate spectrum mentioned above, LP16 have proposed polarization frequency analysis (PFA) and polarization spatial analysis (PSA) to measure the properties of the MHD turbulence cascade. The former considers the change in the variance of polarization intensity at the same LOS as a function of the square of the wavelength $\lambda^2$. In contrast, the latter uses the spatial correlations of the polarization intensity at the same wavelength as a function of the spatial separation. Based on synthetic observations, \cite{Zhang2016ApJ} numerically confirmed that the PFA technique can recover the scaling slope of the underlying turbulent magnetic field under the condition of the Faraday rotation dominated by the mean magnetic field. However, early numerical testing ignored the influence of density distribution on the PFA technique. In addition, recent studies demonstrated that the PFA can also determine the magnetization of the Galactic ISM (\citealt{Guo2024ApJ}).

In this work, we explore how the density distribution coupling magnetic fields affect the measurement of the PFA technique, whether the PFA is subject to the mean magnetic field-dominated Faraday depolarization, and whether the PFA can be applied to complex and diverse interstellar turbulent environments. This paper is organized as follows. Section \ref{Sect_theory} introduces the basis of synchrotron polarization radiation and the statistical methods. Section \ref{Sect_data} describes the characteristics of simulation data cubes and the methods to generate synthetic 3D data. We present our numerical results in Section \ref{Sect_results}, followed by discussion and summary in Sections \ref{Sect_discussion} and \ref{Sect_summary}, respectively.

\section{THEORETICAL DESCRIPTION}\label{Sect_theory}
\subsection{Basic of Synchrotron Polarization Emission}\label{Sect_theory_Synchro-polar}

The complex polarization vector can describe the linear polarization feature of the fluctuations of the synchrotron polarization,
\begin{equation}
P(\bm{X},\lambda^2)=\int_{0}^{L} dz P_i(\bm{X},z)e^{2i\lambda^2\mathrm{RM}(\bm{X},z)}
\label{eq_PIV}
\end{equation}
as a function of the spatial coordinate $\bm{X}$ and the square of wavelength $\lambda^2$, describing a process in which the polarization emissions produced at the source propagate to the observer along the LOS and suffer from Faraday rotation. Here, $P_i(\bm{X},z)$ and $L$ denote the polarization intensity density and the spatial scales of emission transfer, respectively. As is well known, the Faraday rotation measure (RM) is given by
\begin{equation}
\mathrm{RM}(\bm{X},z)=0.81 \int_{0}^{z} \rho_e(\bm{X},z')B_\parallel (\bm{X},z')dz' \rm{rad}~\rm{m^{-2}},
\label{eq_RM}
\end{equation}
where $\rho_e$ and $B_\parallel$ are the density of thermal electrons and the component of the magnetic field along the LOS, respectively. The observed polarization angle is $\phi = \phi_0 + \lambda^2 \rm{RM}$, where $\phi_0$ is the intrinsic polarization angle. 

According to Equation (\ref{eq_PIV}), we can define $P(\bm{X},\lambda^2)=Q+iU$, where $Q$ and $U$ are the observable Stokes parameters $Q$ and $U$. In general, we have $P(\bm{X})=Q_0+iU_0$ in the case of negligible Faraday rotation effect, where $Q_0$ and $U_0$ are called the intrinsic Stokes parameters corresponding to the emitting source zone (refer \citealt{WaelkensMN2009} for calculation formulae). After involving Faraday rotation, we have 
\begin{equation}
Q(\textbf{X}, \lambda^2) = Q_0 \cos{2 \phi} +U_0 \sin{2 \phi}
\label{eq_Stokes_Q}
\end{equation}
and
\begin{equation}
U(\textbf{X}, \lambda^2) = U_0 \cos{2 \phi} -Q_0 \sin{2 \phi}, 
\label{eq_Stokes_U}
\end{equation}
from which we obtain the synchrotron polarization intensity of $P=\sqrt{Q^2+U^2}$.

\subsection{Statistical Description of Polarization Intensity}\label{Sect_Statistical_Description} 
To reveal the properties of MHD turbulence, we use the correlation function of polarization intensity $P$, defined by
\begin{equation}
\mathrm{CF}(\bm{R}) =\langle P(\bm{X_1})P(\bm{X_2}) \rangle,
\label{eq_P_CF}
\end{equation}
where $\bm{R}=\bm{X_2}-\bm{X_1}$ is a separation vector in the plane of the sky. Based on Equation (\ref{eq_PIV}), the correlations of $P$ can be written as
\begin{equation}
\begin{aligned}
&\langle P(\bm{X_1}, \lambda_1^2)P^*(\bm{X_2}, \lambda_2^2) \rangle=\int_{0}^{L}dz_1 \int_{0}^{L}dz_2 \\
& \langle P_i(\bm{X_1},z_1)P_i^*(\bm{X_2},z_2)e^{2i(\lambda_1^2 \rm{RM}_1 - \lambda_2^2 \rm{RM}_2)}.
\label{eq_CF_p2}
\end{aligned}
\end{equation}
Combining the definition of $P \equiv Q+iU$, at any fixed wavelength, Equation (\ref{eq_CF_p2}) can be separated into real and imaginary parts
\begin{equation}
\begin{aligned}
\langle P(\bm{X_1})P(\bm{X_2}) \rangle=&\langle Q(\bm{X_1})Q(\bm{X_2})+U(\bm{X_1})U(\bm{X_2})\rangle +\\
& i \langle U(\bm{X_1})Q(\bm{X_2})-Q(\bm{X_1})U(\bm{X_2})\rangle, 
\label{eq_CF_QU}
\end{aligned}
\end{equation}
which is called two-point statistics (LP16).

In the extreme case of $\textbf{R}=0$ (along a fixed LOS), the spatial correlation of $P$ degenerates into variance, called a one-point statistic. Considering multiple frequency points, the definition of the polarization intensity variance is given by 
\begin{equation}
\begin{aligned}
&\langle P(\lambda^2)P^*(\lambda^2) \rangle=\langle P^2(\lambda^2)\rangle=\\
&\int_{0}^{L}dz_1 \int_{0}^{L}dz_2 \times e^{2i\overline{\Phi} \lambda^2 (z_1-z_2)}\langle P_i(z_1)P_i^*(z_2)e^{2i\lambda^2 [\mathrm{RM}(z_1) -\mathrm{RM}(z_2) ]}\rangle,
\label{eq_CF_QU}
\end{aligned}
\end{equation}
where $\overline{\Phi}=\langle \rho_e(z)B_{\rm \parallel }\rangle$ characterizes the contribution of the mean RM density.

LP16 predicted that the variance of polarization intensity as a function of $\lambda^2$ has different power-law behaviours for different wavelengths. In the long wavelength range (see LP16 for the definition of long wavelength range), the variance of polarization intensity follows 
\begin{equation}
\langle P^2(\lambda^2) \rangle \propto \lambda^{-2-2m},~ \overline{\Phi}>\sigma_\Phi
\label{eq_strongFR_mean}
\end{equation}
and 
\begin{equation}
\langle P^2(\lambda^2) \rangle \propto \lambda^{-2},~ \overline{\Phi}<\sigma_\Phi,
\label{eq_strongFR_fluctuation}
\end{equation}
where $\sigma_\Phi$ and $m$ represent the root mean square of the RM density fluctuation, respectively. The former characterizes the contribution of the turbulent component of RM density and the latter depicts the scaling index of magnetic turbulence ($m=2/3$ for Kolmogorov spectrum). As conveyed by Equation (\ref{eq_strongFR_mean}), the polarization variance can reveal the scaling slope of magnetic turbulence in the case of Faraday rotation dominated by the mean field. However, Equation (\ref{eq_strongFR_fluctuation}) masks the scaling slope of magnetic fluctuations. In this regard, we can also recover the information of magnetic turbulence by the derivative of $\lambda^2 \langle P^2\rangle$ with respect to $\lambda^2$ (LP16), i.e.,
\begin{equation}
{\rm d}(\lambda^2 \langle P^2 \rangle )/{\rm d} \lambda^2 \propto \lambda^{-2-2m}.
\label{eq_derivation}
\end{equation}

In the short wavelength range (see LP16 for the definition of short wavelength range), corresponding to a weak Faraday rotation, polarization variance follows 
\begin{equation}
\langle P^2(\lambda^2) \rangle \propto \lambda^{-2+2m},~ \overline{\Phi}>\sigma_\Phi,
\label{eq_weakFR_mean}
\end{equation}
and
\begin{equation}
\langle P^2(\lambda^2) \rangle \propto \lambda^\frac{-2+2m}{1-\widetilde{m_{\phi}}/2},~ \overline{\Phi}<\sigma_\Phi,
\label{eq_weakFR_fluctuation}
\end{equation}
where $\widetilde{m_{\phi}} = \mathrm{min} (m_{\phi},1)$ with $m_{\phi}$ being the correlation index for the Faraday RM density.

\section{Method of Obtaining Data Cubes}
\label{Sect_data}
To simulate the diffuse ISM with the magnetic field $B$, velocity $v$, and density $\rho$, we use data cubes (see Table \ref{table_data}) generated by the third-order accurate hybrid nonoscillatory code (\citealt{CL03MN}) by solving the ideal isothermal MHD equations with the periodic boundary condition. When generating data cubes with a resolution of $512^3$, we set the fixed parameters including initial injection wavenumber of $k\simeq 2.5$, plasma density of $\rho_p=1.0$, and the box length of $2\pi$, and the variable parameters including the mean magnetic field strength $B_0$ and the gas pressure $P_0$ (related to the sound speed $c_{\rm s}$ by $P_0=\rho_p c_{\rm s}^2$). The total magnetic field $B$ is composed of a background field $B_0$ (along the $x$-axis direction) and a fluctuation field $\delta B$, i.e., $B=B_0+\delta B$. After the evolution reaches a steady state, we calculate the Alfv\'enic Mach number $M_\mathrm{A} = \left\langle\frac{| \bm{v}|}{V_\mathrm{A}}\right\rangle$, the sonic Mach number $M_\mathrm{s}=\left\langle\frac{| \bm{v}|}{c_\mathrm{s}}\right\rangle$, and the plasma $\beta = 2M_{\rm A}^2/M_{\rm s}^2$ to characterize the properties of MHD turbulence, where $\bm{v}$ and $V_\mathrm{A}$ are the velocity of turbulence and Alfv\'enic velocity, respectively. 

Based on the original simulation density data cube $\rho_{\rm ori} (\bm{r})$, we generate a modified density data cube $\rho_{\rm mod} (\bm{r})$ by the Fourier transformation, where $\bm{r}$ is the position vector. In the Fourier space, the density distributions of both $\widetilde{\rho}_{\rm ori} (\bm{k})\propto\bm{k}^{-\alpha_1}$ and $\widetilde{\rho}_{\rm mod} (\bm{k})\propto\bm{k}^{-\alpha_2}$ are associated with the following relationship: 
\begin{equation}
\widetilde{\rho}_{\rm mod}(\bm{k}) = \widetilde{\rho}_{\rm ori}(\bm{k})\bm{k}^{-\alpha_2+\alpha_1}, 
\label{eq_modify}
\end{equation}
where $\bm{k}$ denotes the wavevector. In addition, we can synthesize data cube $\rho_{\rm syn}$ with a normal distribution at an arbitrary spectral index $\alpha$. We first generate density distribution of $\widetilde{\rho}_{\rm syn}(\bm{k}) \propto k^{-\alpha} $ in Fourier space, and then transform it into the real space by the transformation relation (\citealt{Zhang2016ApJ})
\begin{equation}
\rho_{\rm syn}(\bm{r}) = \sum_{k_{\rm min} \leq |\bm{k}| \leq k_{\rm max}} \widetilde{\rho}_{\rm syn}(\bm{k}) e^{i\bm{k} \cdot \bm{r}},
\label{eq_Fourier}
\end{equation}
with the Fourier coefficient $\widetilde{\rho}_{\rm syn}(\bm{k})=|\widetilde{\rho}_{\rm syn}(\bm{k})|e^{i\xi}$, where the phase factor $\xi$ obeys a random normal distribution.

To study the influence of the direction of the mean magnetic field on the polarization variance, we rotate data cubes using the Euler rotation algorithm (\citealt{Parent2012}). We assume the original data cube rotated by the angle $\varphi_{m=x,y,z}$ along the $x$, $y$, and $z$ axes, respectively. Using the rotation matrix of $\hat F = \hat F_x \hat F_y \hat F_z$ (see \citealt{Malik2023MNRAS, Zhang2023MNRAS} for more details), we can obtain postprocessed data cubes: the rotated density by $\rho_{\rm rot}(\bm {r})=\rho_{\rm ori}(\hat F^{-1}\bm {r}) $, and the rotated magnetic field by $\bm{B}_{\rm rot}(\bm {r})=\hat F \bm{B}_{\rm ori}(\hat F^{-1}\bm {r})$.

\begin{deluxetable}{cccccccc}
\caption{Parameters of MHD Turbulence Data Cubes.}
\label{table_data}
\tabletypesize{\scriptsize}
\setlength{\tabcolsep}{3mm}
\tablehead{
\colhead{Data} & \colhead{$B_0$}
   & \colhead{$M_\mathrm{A}$}
   & \colhead{$M_\mathrm{s}$}
   & \colhead{$\beta$}
   & \colhead{$P_0$}
   & \colhead{Skewness}
   & \colhead{Kurtosis}
}
\startdata
D1      & 1.0  & 0.65  & 0.48  & 3.67  &2     &0.03  &-0.24  \\
D2      & 1.0  & 0.58  & 3.17  & 0.07  &0.05  &4.25  &37.03  \\
D3      & 0.1  & 1.72  & 0.45  & 29.22 &2     &-0.37 &0.52   \\
D4      & 0.1  & 1.51  & 4.32  & 0.24  &0.025 &10.31 &319.43\\
\enddata
\begin{tablenotes}
        \footnotesize
        \item[1] \textbf{Notes.} $B_0$--mean strength of magnetic field; $M_\mathrm{A}$--Alfv\'enic Mach number; $M_\mathrm{s}$--sonic Mach number; $\beta=2M_{\rm A}^2/M_{\rm s}^2$--plasma parameter; $P_0$--gas pressure; skewness and kurtosis--numerical statistics of density skewness and kurtosis.
      \end{tablenotes}
\end{deluxetable}

\section{Numerical Results}
\label{Sect_results}

To generate the simulated observations, we take the typical values of the Galactic ISM to parameterize dimensionless quantities: the magnetic field strength of $B=5\ \mu \rm{G}$, the spatial scale of synchrotron emission of $L =500\ \rm{pc}$, the power-law distribution of the relativistic electrons of $N(\epsilon)\propto\epsilon^{-2.5}$, and the thermal electron density of $\rho_e = 0.05\ \rm{cm^{-3}}$. Here, we consider the thermal electron density $\rho_e$ being proportional to the plasma density $\rho_p$.

\subsection{Basic Properties of Data Cubes of Simulations}

\begin{figure*}
\centering  
\vspace{-0.35cm} 
\subfigtopskip=5pt 
\subfigbottomskip=2pt 
\subfigcapskip=-5pt 
\subfigure{
	\label{level.sub.1}
	\includegraphics[width=0.85\linewidth]{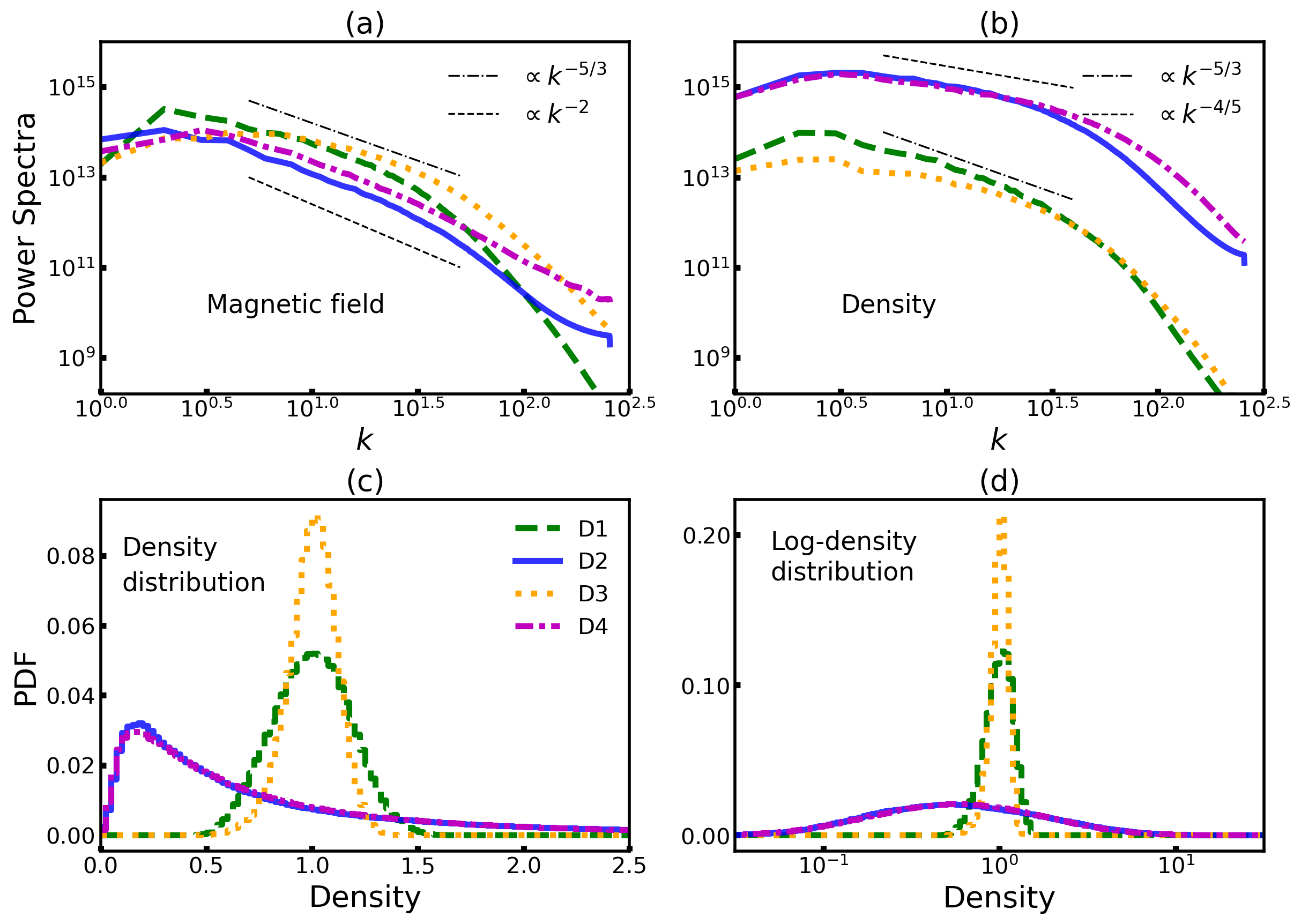}}
\quad 
\caption{The power spectra of magnetic field and density (panels (a) and (b)), and the probability density distributions of plasma density (panels (c) and (d)) arising from different turbulence regimes (see Table \ref{table_data}). 
}
\label{fig: PDF_density-ori}
\end{figure*}

Before exploring synchrotron polarization variance, we first understand the characteristics of data cubes listed in Table \ref{table_data}. Figure \ref{fig: PDF_density-ori} shows the power spectra of magnetic fields (panel (a)) and densities (panel (b)) and the probability distribution functions (PDFs) of densities (panels (c) and (d)). As shown in panel (a), magnetic fields follow the Kolmogorov spectrum of $E_{\rm B} \propto k^{-5/3}$ for subsonic turbulence (see D1 and D3), and are close to $E_{\rm B} \propto k^{-2}$ for supersonic turbulence with the presence of shock waves (see D2 and D4). For the power spectrum of density (see panel (b)), it follows shallower spectra of $E_{\rm \rho} \propto k^{-4/5}$ in the case of supersonic turbulence (D2 and D4), caused by density accumulation at small-scale structures due to the interaction of shock waves. In the case of subsonic turbulence, spectral distribution is close to Kolmogorov-type turbulence of $E_{\rm \rho}\propto k^{-5/3}$ for sub-Alfv\'enic turbulence (see D1). However, we see that the spectrum of D3 presents a shallower spectrum of $E_{\rm \rho}\propto k^{-4/5}$, which may be due to the weak compressibility with a plasma parameter of $\beta=2M_{\rm A}^2/M_{\rm s}^2 \simeq 29.22$.

Panels (c) and (d) show the probability density distributions of plasma density in terms of the linear $\rho_p$ and the logarithmic $\rho_p$, respectively. From panel (c), we see that the linear $\rho_p$ for D1 and D3 follows normal distributions, in agreement with the close-zero values of kurtosis and skewness listed in Table \ref{table_data}. In contrast, the linear $\rho_p$ for D2 and D4 deviates from the normal distribution, with larger values of kurtosis and skewness due to the presence of shock waves in supersonic turbulence. When plotting the PDF of density in the logarithmic $\rho$, we see that $\log \rho_p$ satisfies the normal distribution for each case, which is consistent with the theoretical expectations (\citealt{Passot1998PhRvE}).

\subsection{Application of Polarization Variance to Different Turbulence Regimes }

\begin{figure*}
\centering  
\vspace{-0.35cm} 
\subfigtopskip=5pt 
\subfigbottomskip=2pt 
\subfigcapskip=-5pt 
\subfigure{
	\label{level.sub.1}
	\includegraphics[width=0.85\linewidth]{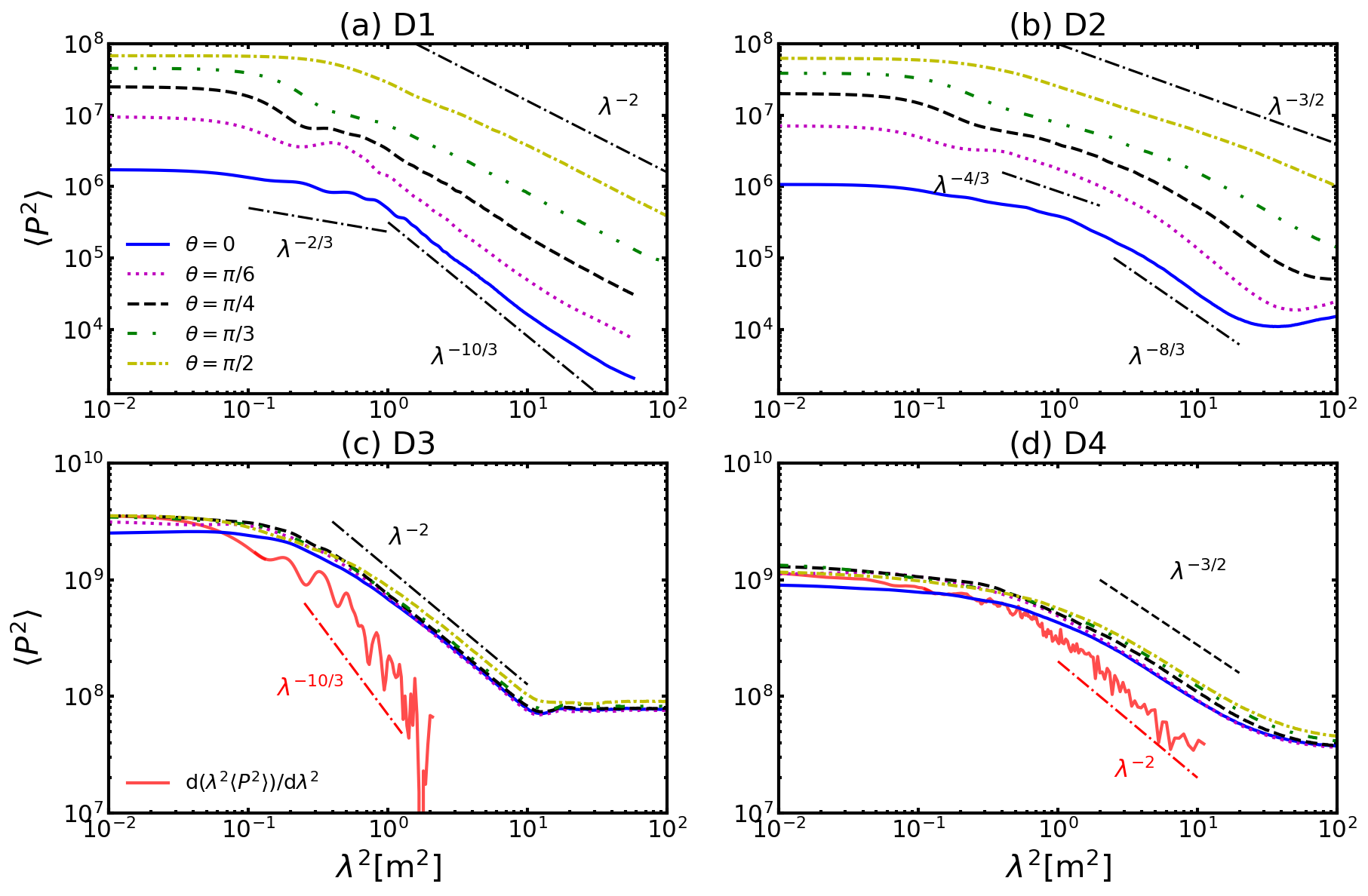}}
\quad 
   \caption{Polarization variance $\langle P^2 \rangle$ as a function of the wavelength squared $\lambda^2$ in four turbulence regimes. The distributions of $d(\lambda^2\langle P^2 \rangle)/d\lambda^2$ are plotted in the lower panels D3 and D4 for the case of $\theta=\pi/2$, where $\theta$ represents the angle between the LOS and the mean magnetic field.}
   \label{fig: P2_lambda_four-cases}  
\end{figure*}

In this section, we apply the polarization variance technique to four turbulence regimes listed in Table \ref{table_data}. By changing the angle $\theta$ between the LOS and the mean magnetic field, resulting in the value change in $\overline{\Phi}=\langle \rho_e(z)B_{\rm \parallel }\rangle$ and its root mean square $\sigma_\Phi$, we explore how the rotation measure dominated by the mean or random fields to affect the relationship between $\langle P^2 \rangle$ and $\lambda^2$. Figure \ref{fig: P2_lambda_four-cases} shows the polarization variance $\langle P^2 \rangle$ as a function of observational wavelength squared $\lambda^2$. As shown in panels (a)-(d), the distribution of polarization variance $\langle P^2 \rangle$ with the increase of wavelength squared $\lambda^2$ can be divided into four stages: (1) the first plateau stage due to negligible Faraday depolarization; (2) shallow power-law stage due to weak Faraday depolarization (see Equations (\ref{eq_weakFR_mean}) and (\ref{eq_weakFR_fluctuation})); (3) steep power-law stage due to strong Faraday depolarization (see Equations (\ref{eq_strongFR_mean}) and (\ref{eq_strongFR_fluctuation})); and (4) the second approximate plateau stage caused by the limited numerical resolution (see \citealt{Zhang2016ApJ} for numerical confirmation). Below, we focus on the shallow and steep power-law stages to reveal the scaling slope of turbulence. 
 
In the case of sub-Alfv\'enic and subsonic turbulence (panel (a)), we have $\langle P^2\rangle \propto \lambda^{-2+2m} = \lambda^{-2/3}$ for the shallow power-law stage, and $\langle P^2\rangle \propto \lambda^{-2-2m} = \lambda^{-10/3}$ ($\theta=0$) as well as $\langle P^2\rangle \propto \lambda^{-2}$ ($\theta=\pi/2$) for the steep power-law stage. When the measured slope is $\langle P^2\rangle \propto \lambda^{-\delta}$, we have 
\begin{equation}
m=(\delta-2)/2,\ {\rm or} \ m=(2-\delta)/2,\label{eq_pv_tur} 
\end{equation}
corresponding to the steep power-law stage or shallow power-law stage. We want to emphasize that $m$ is related to the scaling slope of MHD turbulence: 
\begin{equation}
E_{\rm 3D} \propto k^{-(m+1)},\ {\rm and} \ E_{\rm 2D} \propto k^{-(m+2)}. \label{eq_pv_tur_relation} 
\end{equation}
Note that from $\theta = \pi/2$ (the LOS $\perp \bm{B}$) to $\theta = 0$ (the LOS $\parallel \bm{B}$), the ratio of $\overline{\Phi}/\sigma_\Phi$ changing from 0.06 to 12.88 corresponds to the cases from random RM-dominant regime ($\overline{\Phi}<\sigma_\Phi$) to uniform RM-dominant regime ($\overline{\Phi}>\sigma_\Phi$). It demonstrates that in the case of the dominated mean magnetic field, the index of $m=2/3$ related to the polarization variance can reveal the scaling slope $E_{\rm 3D} \propto k^{-5/3}$ of the magnetic field (see D1 of Figure \ref{fig: PDF_density-ori}a).

For sub-Alfv\'enic and supersonic turbulence (panel (b)), we see that the relation of $\langle P^2\rangle$ vs. $\lambda^2$ in the steep power-law stage follows $\langle P^2\rangle \propto \lambda^{-3/2}$ for the RM dominated by the random field, and $\langle P^2\rangle \propto \lambda^{-8/3}$ for the RM dominated by the mean field. According to Equation (\ref{eq_pv_tur}), we obtain $m=1/3$ (for the latter) related to the measured slope of $E_{\rm 3D} \propto k^{-4/3}$, which significantly deviates from the scaling index 2 of the magnetic field (see D2 of Figure \ref{fig: PDF_density-ori}a). We believe that this deviation stems from the compressibility of the shock waves, resulting in a log-normal distribution of density (see D2 of Figure \ref{fig: PDF_density-ori}d). In the next section, we will explore what can be revealed in the case of the strong coupling between density and magnetic field.

In the case of super-Alfv\'enic turbulence (panels (c) and (d)), we have $\langle P^2\rangle \propto \lambda^{-2}$ (panel (c)) and $\langle P^2\rangle \propto \lambda^{-3/2}$ (panel (d)) at each $\theta$, similar to panels (a) and (b)'s, respectively. It is because in two cases, despite the change in angle $\theta$ from $\theta=\pi/2$ to $\theta=0$, the random field still dominates the Faraday depolarization. As for super-Alfv\'enic and subsonic case (panel (c)), we obtain ${\rm d}(\lambda^2 \langle P^2 \rangle )/{\rm d} \lambda^2 \propto \lambda^{-10/3}$ ($m=2/3$) for the steep power-law stage, which recover the scaling index $5/3$ of turbulent magnetic field. As for super-Alfv\'enic and supersonic case (panel (d)), we find ${\rm d}(\lambda^2 \langle P^2 \rangle )/{\rm d} \lambda^2 \propto \lambda^{-2}$ for the steep power-law stage, which reflects the scaling properties of the RM (power spectral index close to 2 of RM not shown here).

\subsection{Determining the Scaling Slope of Rotation Measure by Polarization Variance}\label{Sect_results_Density}

\begin{figure}
\centering  
\vspace{-0.35cm} 
\subfigtopskip=5pt 
\subfigbottomskip=2pt 
\subfigcapskip=-5pt 
\subfigure{
	\label{level.sub.1}
	\includegraphics[width=0.9\linewidth]{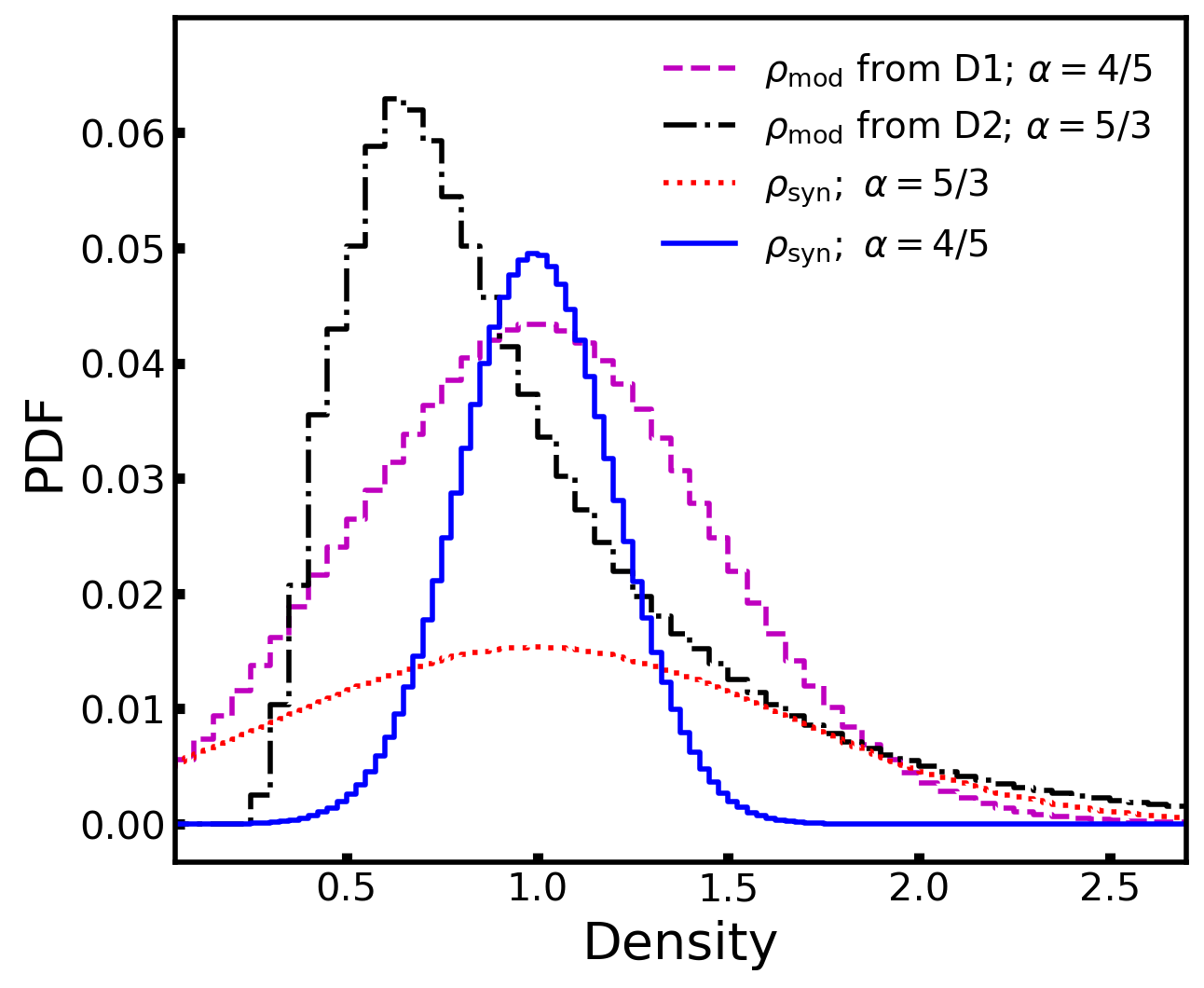}}
\quad 
\caption{The probability density distributions of synthetic ($\rho_{\rm syn}$) and modified ($\rho_{\rm mod}$) densities. $\alpha$ represents the power-law index of density.
}
\label{fig: PDF_syn_mod} 
\end{figure}

To study the effect of density distribution on polarization variance, we first generate different distributions of density as shown in Figure \ref{fig: PDF_syn_mod} by using the method provided in Section \ref{Sect_data}. Based on the density data from D1 ($E_{\rm \rho_{\rm ori}} \simeq k^{-5/3}$) and D2 ($E_{\rm \rho_{\rm ori}} \simeq k^{-4/5}$), we generate the modified density with $E_{\rm \rho_{\rm mod}} = k^{-4/5}$ and $E_{\rm \rho_{\rm mod}} = k^{-5/3}$, respectively. In addition, we also synthesize two data cubes without shock waves: $E_{\rm \rho_{\rm syn}} = k^{-5/3}$ and $E_{\rm \rho_{\rm syn}} = k^{-4/5}$, which maintain the same density spectral indices as D1 and D2, respectively. As shown, the modified and synthesized densities follow a normal distribution of the linear density.

\begin{figure*}
\centering  
\vspace{-0.35cm} 
\subfigtopskip=5pt 
\subfigbottomskip=2pt 
\subfigcapskip=-5pt 
\subfigure{
	\label{level.sub.1}
	\includegraphics[width=0.85\linewidth]{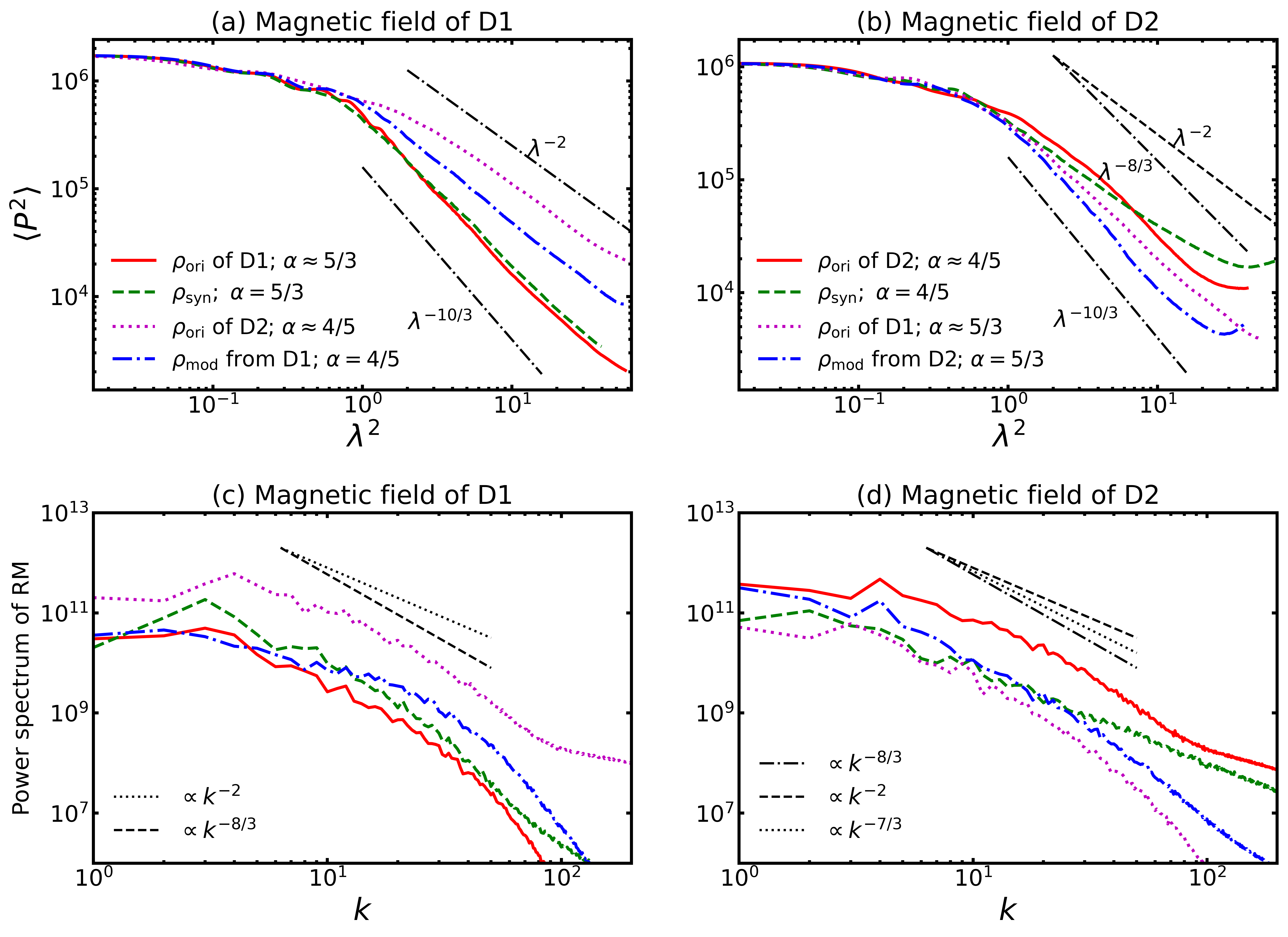}}
\caption{ Upper panels: polarization variance $\langle P^2 \rangle$ vs. the square of wavelength $\lambda^2$, with different density distributions. Lower panels: power spectra of Faraday rotation measure corresponding to the variance distributions plotted by the same line style in the upper panels, respectively. The legends $\rho_{\rm ori}$, $\rho_{\rm mod}$, and $\rho_{\rm syn}$ represent the MHD simulation, modified (from simulation), and synthetic densities, respectively.
}
\label{fig: ori_modify_swap}
\end{figure*}

Focusing on sub-Alfv\'enic turbulence and considering the LOS parallel to the mean magnetic field ($\theta=0$), we study the influence of density on $\langle P^2 \rangle$ in panels (a) and (b) of Figure \ref{fig: ori_modify_swap}. Fixing the D1 magnetic field in panel (a), we compare the relation of $\langle P^2 \rangle$ vs. $\lambda^2$ using four density distributions: (1) original density $\rho_{\rm ori}$ of D1 ($\alpha \approx 5/3$); (2) synthetic density $\rho_{\rm syn}$ with $\alpha = 5/3$; (3) original density $\rho_{\rm ori}$ of D2 ($\alpha \approx 4/5$); and (4) modified density $\rho_{\rm mod}$ from D1 with $\alpha = 4/5$. From this panel, we see that the slope index $10/3$ ($m=2/3$) corresponding to $\rho_{\rm ori}$ of D1 and $\rho_{\rm syn}$, which follow the normal distribution, can reveal the scaling slope of the turbulent magnetic field. Another slope of approximately $2$ corresponding to $\rho_{\rm ori}$ of D2 and $\rho_{\rm mod}$ from D1 cannot reveal the scaling of the turbulent magnetic field. However, we find that the slope of $\langle P^2 \rangle \propto \lambda^{-2}$ ($m=0$ following Equation (\ref{eq_pv_tur})) can reveal the RM scaling slope of $E_{\rm RM} \propto k^{-2}$ as shown in Figure \ref{fig: ori_modify_swap}c.

Similarly, after fixing the D2 magnetic field in panel (b), we show $\langle P^2 \rangle$ vs. $\lambda^2$ by using four density distributions: (1) original density $\rho_{\rm ori}$ of D2 ($\alpha \approx 4/5$); (2) synthetic density $\rho_{\rm syn}$ with $\alpha = 4/5$; (3) original density $\rho_{\rm ori}$ of D1 ($\alpha \approx 5/3$); and (4) modified density $\rho_{\rm mod}$ from D2 with $\alpha = 5/3$. From this panel, we see $\langle P^2 \rangle \propto \lambda^{-10/3}$ for $\rho_{\rm ori}$ of D1 and $\rho_{\rm mod}$ from D2, reflecting the scaling of the turbulent magnetic field due to a weak impact of density, as well as $\langle P^2 \rangle \propto \lambda^{-8/3}$ ($m=1/3$ following Equation (\ref{eq_pv_tur})) for $\rho_{\rm ori}$ of D2, and $\langle P^2 \rangle \propto \lambda^{-2}$ ($m=0$) for $\rho_{\rm syn}$ ($\alpha=4/5$), reflecting the RM scaling slopes: $E_{\rm RM} \propto k^{-7/3}$ and $E_{\rm RM} \propto k^{-2}$ (see also Equation (\ref{eq_pv_tur_relation})), as shown in Figure \ref{fig: ori_modify_swap}d due to a strong density coupling.

As a result, the polarization variance can uncover the power-law characteristics of the 3D magnetic field when the density spatial distribution and its cascade power law do not affect the magnetic field. On the contrary, polarization variance can reflect the power-law distribution of RM.

\subsection{Application of Polarization Variance to Simulated Interferometric Observations}

\begin{figure*}
\centering  
\vspace{-0.35cm} 
\subfigtopskip=5pt 
\subfigbottomskip=2pt 
\subfigcapskip=-5pt 
\subfigure{
	\label{level.sub.2}
	\includegraphics[width=0.85\linewidth]{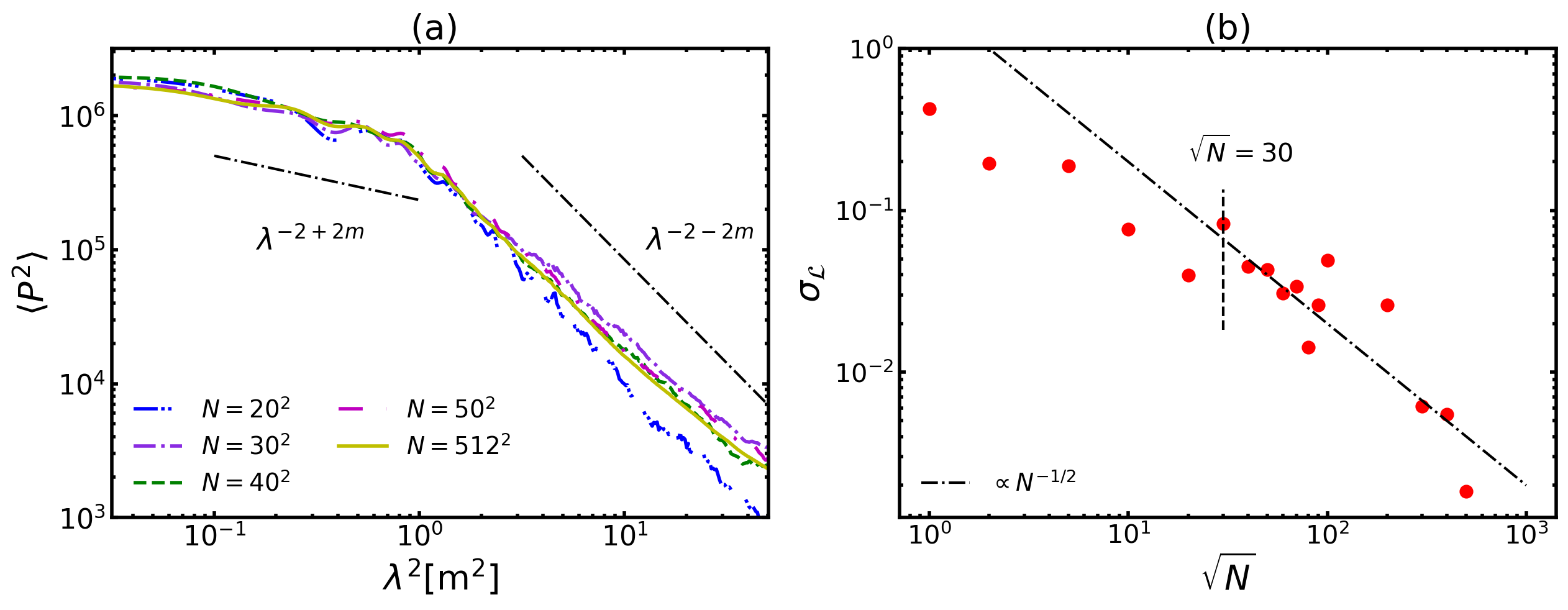}}
\caption{The polarization variance $\langle P^2 \rangle$ vs. the square of wavelength $\lambda^2$ at the randomly selected spatial frequencies (panel (a)) and the error dispersion of polarization variance $\sigma_\mathcal{L}$ vs. the square root of the spatial frequencies (panel (b)). The vertically dashed line represents approximately the spatial frequencies needed for recovering the scaling slope of the turbulent magnetic field or the Faraday rotation measure.
}
\label{fig: b1p2_erro}
\end{figure*}

In the previous sections, our studies demonstrate that the PFA can reveal the scaling slopes of the magnetic field and RM from simulated observations. Here, we study the influence of spatial frequency on revealing the scaling slopes. Following \cite{Zhang2018ApJ}, we consider an observation with the spatial frequencies $N$. By randomly selecting frequencies $N$, we attempt to reconstruct the relation of $\langle P \rangle_{\sqrt{N} \times \sqrt{N}}^2$ vs. $\lambda^2$ with $N \leq 512^2$. As plotted in Figure \ref{fig: b1p2_erro}, we see that with increasing the frequencies, the relation of $\langle P^2 \rangle \propto \lambda^{-2-2m}$ is getting closer to the result of full-frequency sampling ($N=512^2$). According to our simulations, we find that the result corresponding to $N=30^2$ is enough for obtaining the scaling slope of the MHD turbulence.

To quantify the influence of the spatial frequency on polarization variance, we define the measurement error dispersion as $\sigma_\mathcal{L} = \sqrt{(\mathcal{L}-\langle\mathcal{L}\rangle)^2}$, where $\mathcal{L}(\lambda^2) = \langle P\rangle _{\sqrt{N} \times \sqrt{N}}^2 -\langle P\rangle_{512 \times 512}^2$ is the error of measurement, normalized in its mean value. The correlation of $\sqrt{N}$ and the error dispersion $\sigma_\mathcal{L}$ is shown in Figure \ref{fig: b1p2_erro}b, from which we have approximately power law of $\sigma_\mathcal{L} \propto N^{-1/2}$. Therefore, with increasing the spatial frequency, the error dispersion of the measurement decreases.

\section{Discussion}\label{Sect_discussion}
Based on synthetic polarization observation of MHD turbulence simulation data, we tested the practicability of the PFA technique applied to different turbulence regions to obtain the power-law properties of MHD turbulence. The emphasis is on the nature of MHD turbulence that can be revealed in the case of strong or weak coupling between density and magnetic field. With fixing spatial resolution in units of 1/512 pixel and spectral resolution in units of a logarithmic step of 4 MHz, we demonstrated that the finite spectral and spatial resolution are sufficient for revealing the scaling slope of MHD turbulence.

We successfully recovered the scaling slope of the turbulent magnetic field using the polarization variance $\langle P^2 \rangle$ and its derivative ${\rm d}(\lambda^2 \langle P^2 \rangle )/{\rm d} \lambda^2$ in the different turbulence regimes. The former is appropriate for the case where the mean field dominates the RM, and the latter for another case where the random field dominates the RM. For the application of ${\rm d}(\lambda^2 \langle P^2 \rangle )/{\rm d} \lambda^2$ vs. $\lambda^2$, we indeed need higher resolution to obtain smooth enough curves. As demonstrated by \cite{Zhang2016ApJ}, the higher spatial resolution can effectively extend the power-law range of the measured turbulence cascade. In addition, the presence of unavoidable noise will cause the spectrum to bend upwards at the low-frequency regime, shortening the measurable power-law range (see \citealt{Zhang2016ApJ} for $\langle P^2 \rangle$ vs. $\lambda^2$).

Our work studied how density coupling affects the application of the PFA technique. In the case of strong coupling, we cannot obtain the scaling slope of the turbulent magnetic field directly, but the scaling slope of the RM. After using the traditional dispersion measure (called the DM method) to determine density information, we will be able to decouple the magnetic field and density to get the properties of the magnetic field. For instance, in combination with RM and DM, \cite{XuHu2021ApJ} determined the magnetization of the ISM by the anisotropy of the structure function. It should be noted that we cannot directly obtain the magnetic field spectrum from the information of RM and DM.

The traditional method of obtaining the scaling slope of interstellar MHD turbulence is to use the power spectrum. Since this method is subject to the influence of Faraday depolarization, one can only get the power law of the magnetic field perpendicular to the LOS in the high-frequency regime (e.g., \citealt{Zhang2018ApJ}), and cannot determine the power-law characteristics of the 3D magnetic field in the low-frequency regime. In this regard, \cite{Zhang2023MNRAS} attempted to extend the measurable slope range of MHD turbulence by introducing a cross correlation between Stokes parameters $Q$ and $U$. Very recently, the synchrotron polarization gradient analysis has been proposed to measure the spectral properties of MHD turbulence (\citealt{Zhang2025}), which avoids being affected by Faraday depolarization. Similarly, the advantage of the PFA technique is that it avoids the effects of Faraday depolarization. 

Although we use the Galactic ISM from simulation data as a testing platform for the PFA in this work, we expect that the PFA technique can study the magnetic field in different galactic environments. For example, similar to the PSA technique, the PFA should be able to understand the nature of nearby galaxies (e.g., \citealt{Liu2023} for application in the polarization gradient technique). For high-redshift galaxies, the current telescope's spatial resolution, however, is still insufficient. When testing our polarization technique, we consider that synchrotron polarized signals originating from the halo region propagate toward the Galactic plane (e.g., \citealt{Zhang2018ApJ, Wang2021MNRAS}), which is similar to the case that synchrotron polarized signals originating from the external galaxies propagate toward the Milky Way.

We would like to mention that the RM synthesis, which is a classical technique for studying magnetic fields using polarization data, was first proposed by \cite{Burn1966MNRAS} and later improved by \cite{Sokoloff1998MNRAS} and \cite{Brentjens2005A&A}. The application of this method provides important insights into the LOS regular magnetic field component of the Milky Way and neighboring galaxies (e.g., \citealt{Hill2017MNRAS}). However, the RM synthesis technique has a significant limitation in dealing with turbulent magnetic fields. For instance, it cannot reflect the strength and direction of the turbulent field along the LOS (see \citealt{Ferriere2016JPhCS}). In addition, it cannot decouple the density from the magnetic field and reveal the scaling properties of the turbulent magnetic field. Compared with the RM synthesis, the PFA can reveal the scaling slope of the turbulent magnetic field and the RM and can decouple the density from magnetic fields. As a result, our work will effectively complement the RM synthesis techniques, providing significant implications for understanding the nature of ISM.

\section{Summary} \label{Sect_summary}
By extending the study of \cite{Zhang2016ApJ}, our current work demonstrated the feasibility of the PFA technique for revealing the nature of MHD turbulence in different turbulence environments. The main findings of this paper are briefly summarized as follows.

\begin{enumerate} 
\item Spectral properties and probability distribution of density present a significant impact on the application of the PFA technique. 

\item In the case of weak coupling between the density and magnetic field (i.e., subsonic turbulence), the relationships of $\langle P^2 \rangle$ vs. $\lambda^2$ and ${\rm d}(\lambda^2 \langle P^2 \rangle )/{\rm d} \lambda^2$ vs. $\lambda^2$ can recover the scaling slope of the turbulent magnetic fields for the RM dominated by the mean and random fields, respectively.

\item In the cases of a strong coupling (i.e., supersonic turbulence), the relationships of $\langle P^2 \rangle$ vs. $\lambda^2$ and ${\rm d}(\lambda^2 \langle P^2 \rangle )/{\rm d} \lambda^2$ vs. $\lambda^2$ can uncover the scaling slope of the RM for the RM dominated by the mean and random fields, respectively.

\item A finite number of antenna arrays can recover the scaling slope of the turbulent magnetic field. It is of great significance to apply a large number of interferometric data to obtain turbulence properties, such as from the Low-Frequency Array for Radio Astronomy and the Square Kilometre Array.

\end{enumerate} 

\begin{acknowledgments}
We thank the anonymous referee for valuable comments that significantly improved the quality of the paper. J.F.Z. acknowledges the support from the National Natural Science Foundation of China (grant No. 12473046) and the Hunan Natural Science Foundation for Distinguished Young Scholars (No. 2023JJ10039). A.L. acknowledges the support of 1091 NSF grants AST 2307840.
\end{acknowledgments}
\vspace{5mm}

\bibliography{sample631}{}
\bibliographystyle{aasjournal}

\end{document}